\documentstyle[12pt,a4]{article}

\begin{document}
\thispagestyle{empty}
\date{September 21 1993}

\title{%
The relationship between perturbation theory and direct
calculations of rare-earth transition intensities.
}

\author{%
         {Michael F. Reid}\\
          {\em Department of Physics and Astronomy}\\
          {\em University of Canterbury, Christchurch, New Zealand}\\
         {Gary W. Burdick$^*$}\\
          {\em Department of Physics, Southern College}\\
          {\em Collegedale, TN 37315, U.S.A.}\\
         {H. J. Kooy}\\
          {\em Department of Physics, The University of Hong Kong}\\
          {\em Pokfulam Road, Hong Kong.}
}

\maketitle

\begin{abstract}
We use a simplified calculation to demonstrate the equivalence between
three different methods for calculating transition line strengths. These
calculations demonstrate the complex interplay between spin-orbit and
correlation contributions to two-photon transitions in rare-earth ions.
\end{abstract}

Submitted to Journal of Alloys and Compounds,\\
Proceedings of 20th Rare Earth Research Conference

\par \bigskip \par \noindent
$^*$ Current Address:\\
Department of Chemistry, University of Virginia\\
Charlottesville, VA 22901, U.S.A.

\newpage

\section{Introduction}

In a previous paper we investigated many-body perturbation theory
calculations of two-photon transition intensities in Gd$^{3+}$
\cite{BuRe93}.  We found incompatibilities between the many-body
formalism and the earlier calculations of Judd and Pooler
\cite{JuPo82} and Downer and co-workers \cite{Do89} because
the latter contain ``unlinked'' diagrams.  Removal of these diagrams
seemed to destroy the previous good agreement between theory and
experiment.  Agreement was restored, however, when ``folded''
third-order correlation diagrams involving the Coulomb interaction
within the $4f^7$ configuration were included.  The question that
remained unanswered was how these two apparently incompatible
calculations could yield similar answers.

More recently we performed ``direct'' calculations for Eu$^{2+}$, in
which the actual eigenstates of the $4f^7$ and $4f^65d$ configurations
were used \cite{BuKoRe93}.  That work clarified the interplay between
Coulomb and spin-orbit contributions to two-photon transition
intensities and provided a way to reconcile the different methods of
calculation.

It is the purpose of this paper to explicitly demonstrate this
reconciliation.  A key point is that the ``folded'' diagrams represent
changes in energy from the configuration average.  Thus, if we are only
interested in a restricted range of energy levels, we can use the
``true'' energies instead, in which case the Judd-Pooler calculations
give correct results.  To show this, we will briefly review the
Rayleigh-Schr\"{o}dinger perturbation theory.  We then consider a
simplified four-state system, showing the equivalences between direct,
many-body perturbation, and Judd-Pooler type perturbation calculations.

\section{Perturbation theory}

The many-body perturbation-theory approach used in Ref.\ \cite{BuRe93}
is developed in detail in Chapters 9 and 13 of the book by Lindgren and
Morrison \cite{LiMo82}.  When the formalism is extended to include
atom-photon interactions, the energy denominators are modified by the
inclusion of photon energies \cite{PiKe75}, but the same diagrams occur.
The calculations employ the concept of a ``model space''.  Rather than
solving for the eigenvalues and eigenstates of the full Hamiltonian, an
``effective Hamiltonian'' is constructed and diagonalized within the
model space, and the expectation values of ``effective transition
operators'' are evaluated between the model-space eigenvectors.

Initially, we use Rayleigh-Schr\"{o}dinger perturbation theory on the
level of Chapter 9 of Lindgren and Morrison.  If we only consider
excited configurations of opposite parity to the ground configuration,
the  relevant second- and third-order equations for the two-photon
absorption transition moments are
\begin{eqnarray} \label{eq:second}
  &~&
  \sum_{S}
  \frac{\langle K|\mbox{\bf r}|S \rangle
        \langle S|\mbox{\bf r}|I \rangle}
       {(E_I-E_S+\omega)} ,
\\
 \label{eq:thirdstr}
  &~&
  \sum_{ST}
  \frac{\langle K|\mbox{\bf r}|T \rangle
        \langle T|V      |S \rangle
        \langle S|\mbox{\bf r}|I \rangle}
       {(E_I-E_T+\omega)(E_I-E_S+\omega)} ,
\\
 \label{eq:thirdfol}
 &-& \sum_{JS}
  \frac{\langle K|\mbox{\bf r}|S \rangle
        \langle S|\mbox{\bf r}|J \rangle
        \langle J|V      |I \rangle}
       {(E_J-E_S+\omega)(E_I-E_S+\omega)} .
\end{eqnarray}
In these expressions, $I$, $J$, and $K$ are states within the model
space (4$f^N$) and $S$ and $T$ are states in an excited configuration
({\em e.g.} $4f^{N-1}5d$) outside the model space. The perturbation,
$V$, includes the spin-orbit and Coulomb interactions, $\mbox{\bf r}$ is
the dipole moment operator, and $\omega$ is the photon energy
($\hbar=1$).  If desired, Eq.\ (\ref{eq:thirdfol}), may be Hermitized by
averaging with its conjugate.  However, it should be noted that
model-space calculations are inherently non-Hermitian.

When excited configurations having the same parity as the ground
configuration are included, we have the two additional equations,
\begin{eqnarray}
 \label{eq:thirdeven1}
  &~&
  \sum_{MS}
  \frac{\langle K|\mbox{\bf r}|S \rangle
        \langle S|\mbox{\bf r}|M \rangle
        \langle M|V           |I \rangle}
       {(E_I-E_S+\omega)(E_I-E_M)} ,
\\
 \label{eq:thirdeven2}
  &~&
  \sum_{SM}
  \frac{\langle K|V           |M \rangle
        \langle M|\mbox{\bf r}|S \rangle
        \langle S|\mbox{\bf r}|I \rangle}
       {(E_I-E_M+2\omega)(E_I-E_S+\omega)} ,
\end{eqnarray}
where the $M$ are states in an excited configuration of the same parity
as the ground configuration ({\em e.g.} 4f$^{N-1}$5f). For the
simplified calculations that follow, however, only a single excited
configuration will be considered, and thus contributions from Eqs.\
(\ref{eq:thirdeven1}--\ref{eq:thirdeven2}) will not arise.

     The above equations will be used for each of the three
calculations which follow.  The fundamental differences between these
calculations are not in the equations used, but rather in their
differing definitions of the states $|i \rangle$ and energies $E_i$.
For the direct calculation, eigenstates and eigenvalues of the complete
Hamiltonian will be used, allowing only contributions from Eq.\
(\ref{eq:second}).  In contrast, the many-body calculation uses
zero-order eigenstates, necessitating the use of all relevant
equations.  The Judd-Pooler type perturbation calculation uses the
zero-order eigenstates within the excited configuration (like the
many-body calculation) and uses exact eigenstates within the ground
configuration (like the direct calculation).  Thus, Judd-Pooler type
calculations omit Eq.\ (\ref{eq:thirdfol}), which has the perturbation
$V$ acting within the ground configuration.  In the next section, we
will show that when careful attention is paid to these differences, all
three calculations result in equivalent answers.

\section{Simplified calculation}

We consider a simplified system with four states, $|a\rangle$,
$|b\rangle$, $|c\rangle$, and $|d\rangle$.  The model space consists of
$|a\rangle$ and $|b\rangle$, while $|c\rangle$ and $|d\rangle$ are
excited states.  The Hamiltonian, $H$, is divided into a zero-order
part, $H_0$, and a perturbation, $V$,
\begin{equation} \label{eq:ham}
H = H_0 + V =
   \left(
      \begin{array}{cccc}
        0   &0     &0          &0 \\
        0   &0     &0          &0 \\
        0   &0     &\Delta     &0  \\
        0   &0     &0          &\Delta \\
      \end{array}
   \right)
+
   \left(
      \begin{array}{cccc}
        C_a &Z_1   &0          &0 \\
        Z_1 &C_b   &0          &0 \\
        0   &0     &C_c        &Z_2\\
        0   &0     &Z_2        &C_d\\
      \end{array}
    \right) ,
\end{equation}
where $\Delta$ is the zero-order energy difference between the excited
and model spaces (the Hartree-Fock energy). The $C_i$ represent the
non-spherical part of the Coulomb interaction, and the $Z_j$ represent
the spin-orbit interaction.  The effective Hamiltonian for the model
space is
\begin{equation} \label{eq:heff}
 H_{eff} =
      \left(
      \begin{array}{cc}
        C_a &Z_1 \\
        Z_1 &C_b \\
      \end{array}
    \right) .
\end{equation}
In the general case there could be matrix elements of $V$ ({\em e.g.}
the odd-parity crystal field) connecting $|a\rangle$ and $|b\rangle$ to
$|c\rangle$ and $|d\rangle$.  In that case, Eq.\ (\ref{eq:heff}) would be
a first-order approximation to $H_{eff}$.

If the $Z_j$ are small compared to the $C_i$, we can take the mixing to
first order in the $Z_j$ and ignore the energy shift.  This yields the
following expressions for the $|i'\rangle$ eigenstates of $H$:
\begin{eqnarray}
\label{eq:ap}
|a'\rangle = |a\rangle + |b\rangle \frac{Z_1}{C_a-C_b},
           &~& E_{a'}=C_a ~,\\
\label{eq:bp}
|b'\rangle = |b\rangle + |a\rangle \frac{Z_1}{C_b-C_a},
           &~& E_{b'}=C_b ~,\\
\label{eq:cp}
|c'\rangle = |c\rangle + |d\rangle \frac{Z_2}{C_c-C_d},
           &~& E_{c'}=\Delta+C_c ~,\\
\label{eq:dp}
|d'\rangle = |d\rangle + |c\rangle \frac{Z_2}{C_d-C_c},
           &~& E_{d'}=\Delta+C_d  ~.
\end{eqnarray}
The states $|a'\rangle$ and $|b'\rangle$ are also the eigenstates of
$H_{eff}$.

We take the one-photon interaction matrix to be
\begin{equation}
M = \left(
      \begin{array}{cccc}
        0   &0     &T_1  &0 \\
        0   &0     &0    &T_2\\
        T_1 &0     &0    &0 \\
        0   &T_2   &0    &0 \\
      \end{array}
    \right) ,
\end{equation}
where the $T_k$ represent dipole moments. Two-photon transitions between
$|a'\rangle$ and $|b'\rangle$ are ``spin-forbidden'', as they are
allowed only due to the presence of the ``spin-orbit'' matrix elements,
$Z_j$, in the Hamiltonian.

\subsection{Direct calculation}

In order to perform an ``exact'' or ``direct'' calculation of the
two-photon transition moment for the ``spin-forbidden'' $|a'\rangle$ to
$|b'\rangle$ transition, we use exact eigenstates of $H$, rather than
those of $H_0$.  This makes the third-order terms superfluous, and only
the second-order matrix elements of Eq.\ (\ref{eq:second}) must be
evaluated.

Rather than write out all possible terms, we present only those that are
quadratic in $T_2$ and linear in $Z_1$.
{}From Eq.\ (\ref{eq:second}), we obtain
\begin{equation} \label{eq:direct1}
      \left( \frac{T_2 T_2}{C_a - (\Delta+C_d)+\omega} \right)
      \left( \frac{Z_1}{C_a-C_b} \right) ,
\end{equation}
where the first fraction represents the second-order time dependent
perturbation, and the second fraction gives the proportion of
$|b\rangle$ in $|a'\rangle$.
Expanding Eq.\ (\ref{eq:direct1}) in powers of $1/(-\Delta+\omega)$, we obtain
\begin{equation} \label{eq:direct2}
  \frac{T_2 T_2 Z_1}{C_a-C_b}
    \left(
          \frac{1}{-\Delta+\omega} + \frac{C_d-C_a}{(-\Delta+\omega)^2}
                    + \ldots
    \right) .
\end{equation}

\subsection{Many-body perturbation calculation}

The states in Eqs.\ (\ref{eq:second}--\ref{eq:thirdfol}) represent
configurational states, whereas many-body perturbation theory is
expressed in terms of orbitals.  However, the correspondence between
Eqs.\ (\ref{eq:second}--\ref{eq:thirdfol}) and orbital expressions is
exact (see Chapter 13 of \cite{LiMo82}, \cite{NgNe85}), and for the
purposes of this discussion, we use Eqs.\
(\ref{eq:second}--\ref{eq:thirdfol}) to demonstrate the results of a
many-body calculation.  Our calculation is based on
Rayleigh-Schr\"{o}dinger perturbation theory, so eigenvalues of $H_0$
are used in the energy denominators.  Furthermore, the states in Eqs.\
(\ref{eq:second}--\ref{eq:thirdfol}) are the zero-order states
($|i\rangle$, not $|i'\rangle$). In order to obtain the terms containing
$T_2T_2Z_1$, we place these ``effective transition operators'' between
the eigenstates of $H_{eff}$.  This leads to the factors
${Z_1}/({C_a-C_b})$ in some of the following expressions.  From Eq.\
(\ref{eq:second}) we obtain
\begin{equation}
\label{eq:aq}
\frac{T_2 T_2}{-\Delta+\omega} \left( \frac{Z_1}{C_a-C_b} \right) .
\end{equation}
Restricting $V$ to the spin-orbit interaction in Eq.\ (\ref{eq:thirdstr}),
we only get terms involving $Z_2$.  However, letting $V$ contain the Coulomb
interaction yields
\begin{equation}
\label{eq:bq}
\frac{T_2 C_d T_2}{(-\Delta+\omega)(-\Delta+\omega)}
      \left( \frac{Z_1}{C_a-C_b} \right)  .
\end{equation}
Eq.\ (\ref{eq:thirdfol}) yields
\begin{equation}
\label{eq:cq}
- ~ \frac{T_2 T_2 Z_1}{(-\Delta+\omega)(-\Delta+\omega)}
   ~ = ~
        - ~ \frac{T_2 T_2 Z_1}{(-\Delta+\omega)(-\Delta+\omega)}
          \times \left( \frac{C_a-C_b}{C_a-C_b} \right) ,
\end{equation}
\begin{equation}
\label{eq:dq}
- ~ \frac{T_2 T_2 C_b}{(-\Delta+\omega)(-\Delta+\omega)}
             \left( \frac{Z_1}{C_a-C_b} \right) ,
\end{equation}
from the spin-orbit and Coulomb contributions, respectively.  Adding
these terms together, we get
\begin{equation} \label{eq:pttot}
 \frac{T_2 T_2 Z_1}{C_a-C_b}
   \left(
       \frac{1}{-\Delta+\omega}
            + \frac{  C_d -(C_a-C_b) - C_b}{(-\Delta+\omega)^2}
   \right).
\end{equation}
which is the same as the ``direct'' calculation. Note the cancellation
between Eqs.\ (\ref{eq:cq}) and (\ref{eq:dq}), which  arise
from ``spin-orbit'' and ``Coulomb'' interactions, respectively.

\subsection{Judd-Pooler perturbation calculation}

The original Judd-Pooler \cite{JuPo82} calculation does not include Eq.\
(\ref{eq:thirdfol}) and thus Eqs.\ (\ref{eq:cq}-\ref{eq:dq}) do not
arise. However, the denominators now contain eigenvalues of $H_{eff}$,
and Eq.\ (\ref{eq:second}) yields
\begin{equation} \label{eq:exact1}
      \left( \frac{T_2 T_2}{C_a - \Delta+\omega} \right)
      \left( \frac{Z_1}{C_a-C_b} \right) .
\end{equation}
Coulomb contributions within the excited configuration were considered
to be small in the calculations of Judd and Pooler, and were therefore
neglected.  For completeness, however, we include them here.  Thus, from
Eq.\ (\ref{eq:thirdstr}) we obtain
\begin{equation} \label{eq:exact2}
  \left(
         \frac{T_2 C_d T_2}{(C_a - \Delta+\omega)(C_a - \Delta+\omega)}
  \right)
  \left( \frac{Z_1}{C_a - C_b}    \right) .
\end{equation}
Adding these terms together and expanding in powers of
$1/(-\Delta+\omega)$ we obtain
\begin{equation} \label{eq:exact3}
  \frac{T_2 T_2 Z_1}{C_a-C_b}
    \left(
          \frac{1}{-\Delta+\omega} + \frac{C_d-C_a}{(-\Delta+\omega)^2}
                + \ldots
    \right) ,
\end{equation}
which is the same as the results of the direct calculation (Eq.\
(\ref{eq:direct2})) and the many-body perturbation calculation (Eq.\
(\ref{eq:pttot})).  We have emphasized terms containing $T_2 T_2 Z_1$,
in order to display expressions arising from Eq.\ (\ref{eq:thirdfol}).
In general, when terms involving any combination of matrix elements are
evaluated, with perturbation expansions carried out to $n$-th order,
agreement is exact through the $n-1$ order of $1/(- \Delta+\omega)$.
Thus, to achieve agreement between these three calculations in the third
order of $1/(- \Delta+\omega)$, it would be necessary to consider
fourth-order perturbation terms.

In essence, as explained in Ref. \cite{BuKoRe93}, the Judd-Pooler
calculation has moved the part of $V$ that acts within the model space
into $H_0$.  This means that the matrix element $\langle J|V|I\rangle$
of Eq.\ (\ref{eq:thirdfol}) is equal to zero, since there is no longer
any $V$ separate from $H_0$ acting within the model space.  The
Judd-Pooler calculation is therefore correct, but it must be realized
that different energy denominators should be used for transitions
between different energy levels.

In Judd-Pooler-type calculations (Eq.\ \ref{eq:exact1}), terms
containing the Coulomb matrix elements, $C_i$, arise from using
eigenvalues of $H_{eff}$, which includes effects of the Coulomb
interaction.  In many-body perturbation calculations, by contrast, they
arise only from perturbation expressions containing the Coulomb
interaction explicitly.  Therefore, the classification of contributions
as ``spin-orbit'' or ``correlation'' depends, in part, upon the method
of calculation.

\section{Many-body perturbation theory for Gd$^{3+}$}

For the Gd$^{3+}$ calculation discussed in Ref.\ \cite{BuRe93} the
states in Eqs. (\ref{eq:second}--\ref{eq:thirdeven2}) represent states in
the $4f^7$ and $4f^65d$ {\em configurations}.  Many-body perturbation
theory, however, is expressed in terms of {\em orbitals}.  The
transformation from Eqs.  (\ref{eq:second}--\ref{eq:thirdeven2}) to
expressions corresponding to the diagrams evaluated in Ref.\
\cite{BuRe93} is explained by Lindgren and Morrison \cite{LiMo82}
(Chapter 13, see also
\cite{NgNe85}).

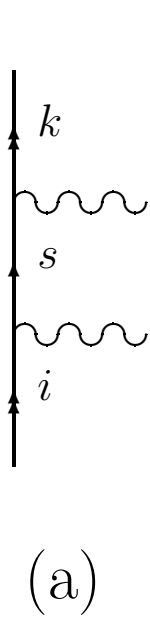
\begin{figure}[h]
 \caption{Second and third-order one-electron
two-photon absorption diagrams. For details of notation see Ref.\
\protect\cite{LiMo82}. A single up-going arrow represents a virtual
orbital, a double arrow a valence orbital. Folded valence lines have a
circle around their arrows. Photons are represented by wavy lines and
the spin-orbit interaction by a triangle.  Diagrams (a), (b),  (c), (e),
and (f) correspond to Eqs.\   (\protect\ref{eq:secord}),
(\protect\ref{eq:zetad}), (\protect\ref{eq:zetaf}),
(\protect\ref{eq:zetaeven1}), and (\protect\ref{eq:zetaeven2}) of the
text, respectively.  Diagram (c) is folded, diagram (d) is unlinked.
     \label{fig:tpa1}}


\begin{picture}( 450.0, 500.0)( -25.0,-275.0)
\thicklines
\put(  50.0, -25.0){ {\Huge (a)}}
\put(  50.0,  25.0){\line(0,1){  50.0}}
\put(  50.0,  55.0){\vector(0,1){0}}
\put(  50.0,  51.0){\vector(0,1){0}}
\put(  55.0,  50.0){ {\Large $i$}}
\multiput(  54.2,  75.0)(  16.7,   0.0){3}{\oval(   8.3,   8.3)[t]}
\multiput(  62.5,  75.0)(  16.7,   0.0){3}{\oval(   8.3,   8.3)[b]}
\put(  50.0,  75.0){\line(0,1){  50.0}}
\put(  50.0, 103.0){\vector(0,1){0}}
\put(  55.0, 100.0){ {\Large $s$}}
\multiput(  54.2, 125.0)(  16.7,   0.0){3}{\oval(   8.3,   8.3)[t]}
\multiput(  62.5, 125.0)(  16.7,   0.0){3}{\oval(   8.3,   8.3)[b]}
\put(  50.0, 125.0){\line(0,1){  50.0}}
\put(  50.0, 155.0){\vector(0,1){0}}
\put(  50.0, 151.0){\vector(0,1){0}}
\put(  55.0, 150.0){ {\Large $k$}}

\put( 250.0, -25.0){ {\Huge (b)}}
\put( 250.0,   0.0){\line(0,1){  50.0}}
\put( 250.0,  30.0){\vector(0,1){0}}
\put( 250.0,  26.0){\vector(0,1){0}}
\put( 255.0,  25.0){ {\Large $i$}}
\multiput( 254.2,  50.0)(  16.7,   0.0){3}{\oval(   8.3,   8.3)[t]}
\multiput( 262.5,  50.0)(  16.7,   0.0){3}{\oval(   8.3,   8.3)[b]}
\put( 250.0,  50.0){\line(0,1){  50.0}}
\put( 250.0,  78.0){\vector(0,1){0}}
\put( 255.0,  75.0){ {\Large $s$}}
\put( 250.0, 100.0){\line(1,0){   3.1}}
\multiput( 259.4, 100.0)(  12.5,   0.0){3}{\line(1,0){   6.3}}
\put( 296.9, 100.0){\line(1,0){   3.1}}
\put( 300.0, 100.0){\line(1,1){  10.0}}
\put( 310.0, 110.0){\line(0,-1){  20.0}}
\put( 310.0,  90.0){\line(-1,1){  10.0}}
\put( 250.0, 100.0){\line(0,1){  50.0}}
\put( 250.0, 128.0){\vector(0,1){0}}
\put( 255.0, 125.0){ {\Large $t$}}
\multiput( 254.2, 150.0)(  16.7,   0.0){3}{\oval(   8.3,   8.3)[t]}
\multiput( 262.5, 150.0)(  16.7,   0.0){3}{\oval(   8.3,   8.3)[b]}
\put( 250.0, 150.0){\line(0,1){  50.0}}
\put( 250.0, 180.0){\vector(0,1){0}}
\put( 250.0, 176.0){\vector(0,1){0}}
\put( 255.0, 175.0){ {\Large $k$}}
\put(  50.0,-250.0){ {\Huge (c)}}
\put( 100.0,-225.0){\line(0,1){  12.5}}
\put( 100.0,-212.5){\line(0,1){  50.0}}
\put( 100.0,-182.5){\vector(0,1){0}}
\put( 100.0,-186.5){\vector(0,1){0}}
\put( 105.0,-187.5){ {\Large $i$}}
\put( 100.0,-162.5){\line(0,1){  12.5}}
\put( 100.0,-150.0){\line(1,0){   3.1}}
\multiput( 109.4,-150.0)(  12.5,   0.0){3}{\line(1,0){   6.3}}
\put( 146.9,-150.0){\line(1,0){   3.1}}
\put( 150.0,-150.0){\line(1,1){  10.0}}
\put( 160.0,-140.0){\line(0,-1){  20.0}}
\put( 160.0,-160.0){\line(-1,1){  10.0}}
\put( 100.0,-150.0){\line(-1,-1){  50.0}}
\put(  70.0,-180.0){\vector(-1,-1){0}}
\put(  74.0,-176.0){\vector(-1,-1){0}}
\put(  75.0,-175.0){\circle{  15.0}}
\put(  65.0,-200.0){ {\Large $j$}}
\multiput(  45.8,-200.0)( -16.7,   0.0){3}{\oval(   8.3,   8.3)[t]}
\multiput(  37.5,-200.0)( -16.7,   0.0){3}{\oval(   8.3,   8.3)[b]}
\put(  50.0,-200.0){\line(0,1){  25.0}}
\put(  50.0,-175.0){\line(0,1){  50.0}}
\put(  50.0,-147.0){\vector(0,1){0}}
\put(  55.0,-150.0){ {\Large $s$}}
\multiput(  45.8,-125.0)( -16.7,   0.0){3}{\oval(   8.3,   8.3)[t]}
\multiput(  37.5,-125.0)( -16.7,   0.0){3}{\oval(   8.3,   8.3)[b]}
\put(  50.0,-125.0){\line(0,1){  50.0}}
\put(  50.0, -95.0){\vector(0,1){0}}
\put(  50.0, -99.0){\vector(0,1){0}}
\put(  55.0,-100.0){ {\Large $k$}}

\put( 250.0,-250.0){ {\Huge (d)}}
\put( 250.0,-225.0){\line(0,1){  50.0}}
\put( 250.0,-195.0){\vector(0,1){0}}
\put( 250.0,-199.0){\vector(0,1){0}}
\put( 255.0,-200.0){ {\Large $i$}}
\multiput( 254.2,-175.0)(  16.7,   0.0){3}{\oval(   8.3,   8.3)[t]}
\multiput( 262.5,-175.0)(  16.7,   0.0){3}{\oval(   8.3,   8.3)[b]}
\put( 250.0,-175.0){\line(0,1){  50.0}}
\put( 250.0,-147.0){\vector(0,1){0}}
\put( 255.0,-150.0){ {\Large $s$}}
\multiput( 254.2,-125.0)(  16.7,   0.0){3}{\oval(   8.3,   8.3)[t]}
\multiput( 262.5,-125.0)(  16.7,   0.0){3}{\oval(   8.3,   8.3)[b]}
\put( 250.0,-125.0){\line(0,1){  50.0}}
\put( 250.0, -95.0){\vector(0,1){0}}
\put( 250.0, -99.0){\vector(0,1){0}}
\put( 255.0,-100.0){ {\Large $k$}}
\put( 315.0,-157.5){ {\Huge $\times$}}
\put( 350.0,-200.0){\line(0,1){  50.0}}
\put( 350.0,-170.0){\vector(0,1){0}}
\put( 350.0,-174.0){\vector(0,1){0}}
\put( 355.0,-175.0){ {\Large $j$}}
\put( 350.0,-150.0){\line(1,0){   3.1}}
\multiput( 359.4,-150.0)(  12.5,   0.0){3}{\line(1,0){   6.3}}
\put( 396.9,-150.0){\line(1,0){   3.1}}
\put( 400.0,-150.0){\line(1,1){  10.0}}
\put( 410.0,-140.0){\line(0,-1){  20.0}}
\put( 410.0,-160.0){\line(-1,1){  10.0}}
\put( 350.0,-150.0){\line(0,1){  50.0}}
\put( 350.0,-120.0){\vector(0,1){0}}
\put( 350.0,-124.0){\vector(0,1){0}}
\put( 355.0,-125.0){ {\Large $l$}}

\end{picture}

\vspace*{1cm}

 \end{figure}

Ignoring diagrams involving core excitations, we have only one
second-order diagram and four third-order diagrams involving the
spin-orbit interaction, $V_{so}$, and the dipole moment, $\mbox{\bf
r}$. These are shown in Fig.\ \ref{fig:tpa1}(a--c, e--f), and have
corresponding algebraic expressions
\begin{eqnarray}\label{eq:secord}
 & &
 a^{\dagger}_k a_i \sum_{s}
 \frac{\langle k |\mbox{\bf r}| s \rangle
       \langle s |\mbox{\bf r}| i \rangle}
      {\varepsilon_i - \varepsilon_s  + \omega} , \\
 \label{eq:zetad}
 & &
 a^{\dagger}_k a_i \sum_{st}
 \frac{\langle k |\mbox{\bf r}| t \rangle
       \langle t |V_{so} | s \rangle
       \langle s |\mbox{\bf r}| i \rangle}
      {(\varepsilon_i - \varepsilon_t + \omega)
       (\varepsilon_i - \varepsilon_s + \omega)} , \\
\label{eq:zetaf}
 &-& a^{\dagger}_k a_i \sum_{js}
 \frac{\langle k |\mbox{\bf r}| s \rangle
       \langle s |\mbox{\bf r}| j \rangle
       \langle j |V_{so} | i \rangle}
      {(\varepsilon_i - \varepsilon_s+ \omega)
       (\varepsilon_j - \varepsilon_s+ \omega)} , \\
\label{eq:zetaeven1}
 & & a^{\dagger}_k a_i \sum_{ms}
 \frac{\langle k |\mbox{\bf r}| s \rangle
       \langle s |\mbox{\bf r}| m \rangle
       \langle m |V_{so} | i \rangle}
      {(\varepsilon_i - \varepsilon_s+ \omega)
       (\varepsilon_i - \varepsilon_m)}, \\
\label{eq:zetaeven2}
 & & a^{\dagger}_k a_i \sum_{ms}
 \frac{\langle k |V_{so}      | m \rangle
       \langle m |\mbox{\bf r}| s \rangle
       \langle s |\mbox{\bf r}| i \rangle}
      {(\varepsilon_i - \varepsilon_m +2 \omega)
       (\varepsilon_i - \varepsilon_s+\omega)} .
\end{eqnarray}
In these expressions, $i$, $j$, and $k$ label valence ($4f$) orbitals in
our $4f^7$ model space, $s$ and $t$ label virtual orbitals outside the
model space ($5d$), and the $\varepsilon$ are orbital energies.
Each term contains a product of a creation operator and an annihilation
operator for $4f$ orbitals. The diagram corresponding to Eq.\
(\ref{eq:zetaf}) (Fig.\ \ref{fig:tpa1}(c)) is commonly referred to as
``folded'' \cite{LiMo82}.

Eq.\ (\ref{eq:secord}) arises directly from Eq.\ (\ref{eq:second}).
Eq.\ (\ref{eq:zetad}) arises from Eq.\ (\ref{eq:thirdstr}), where the
spin-orbit interaction and the dipole moments act on the same electron.
``Unlinked'' expressions (Fig.\ \ref{fig:tpa1}(d)), where the dipole
moments and the spin-orbit operator act on different electrons, arise
from both Eq.\ (\ref{eq:thirdstr}) and Eq.\ (\ref{eq:thirdfol}) and
cancel. This leaves us with the ``folded'' diagram (Fig.\
\ref{fig:tpa1}(c), Eq.\ (\ref{eq:zetaf})) which arises from Eq.\
(\ref{eq:thirdfol}) when the dipole moments and spin-orbit operator act
on the same electron.

It is important not to confuse ``folded'' diagrams with ``core
excitations''. The folding is merely a device to preserve the rules for
calculating the denominators \cite{LiMo82} (especially pp. 304--305).
In Fig.\ \ref{fig:tpa1}(c), for example, an electron is excited from
state $i$ to $j$ to $s$ to $k$, and the diagram does {\em not} imply
that there is initially an electron in state $j$, as is clear from the
operator ordering of Eq.\ (\ref{eq:zetaf}).

Eqs.\ (\ref{eq:zetaeven1}) and (\ref{eq:zetaeven2}) arise from Eqs.\
(\ref{eq:thirdeven1}) and (\ref{eq:thirdeven2}). These have been
overlooked in earlier work \cite{JuPo82,Do89,BuRe93}. Their possible
importance has been recently demonstrated by Smentek-Mielczarek \cite{Sm93}.

Some third-order ``correlation'' diagrams involving the Coulomb
interaction are shown in Fig.\ \ref{fig:tpa2}. The folded diagram, Fig.\
\ref{fig:tpa2}(e), arises from Eq.\ (\ref{eq:thirdfol}), when $V$ is the
Coulomb interaction. It gives a large contribution in the calculations
of Ref.\ \cite{BuRe93}, since it represents perturbations to the
energy-levels of the $f^{7}$ configuration caused by the Coulomb
interaction. An analogous term in our simplified calculation is Eq.\
(\ref{eq:dq}).  As in our simplified calculation, it is possible to take
these effects into account implicitly by using the exact energy levels.
Thus, the agreement between the calculation of Ref.\ \cite{BuRe93} and
the calculation of Judd and Pooler \cite{JuPo82} (in which the
correlation effects are absorbed into the denominators) is not merely
fortuitous, but is a result of doing the same calculation in different
ways.

\section{Conclusion}

Careful application of three different calculation methods (direct
\cite{BuKoRe93}, many-body perturbation \cite{BuRe93}, and Judd-Pooler
\cite{JuPo82,Do89} type perturbation calculations) yields identical
results. Thus, each of these three methods may legitimately be used in
the calculation of transition line strengths.

The Judd-Pooler \cite{JuPo82} calculation, while valid, is
not compatible with Lindgren and Morrison's \cite{LiMo82} formulation of
many-body perturbation theory.  This is because the Judd-Pooler
zero-order Hamiltonian, $H_0$, is different when acting upon ground
configuration states than when acting upon states of the excited
configuration, thus destroying the cancellation of unlinked diagrams
necessary for the many-body perturbation theory.

In contrast, the direct calculations used in Ref. \cite{BuKoRe93} and
illustrated here are compatible with MBPT, as long as we define the
model space to include both the $4f^N$ and the $4f^{N-1}5d$
configurations. In that case it is possible to add contributions from
other excited configurations or from other potential terms, using the
techniques of MBPT.  In fact, we feel that this will be a most
profitable direction for future calculations, particularly for
examination of transition intensities for divalent lanthanides, where
both the Judd-Pooler formalism \cite{Do89} and the more straight-forward
MBPT calculations, using $4f^N$ as the model space, have been shown to be
insufficient \cite{BuKoRe93}.

\subsection*{Acknowledgements}

We thank B.R. Judd  and L. Smentek-Mielczarek for helpful discussions.




\begin{thebibliography}{1}

\bibitem{BuRe93}
G.~W. Burdick and M.~F. Reid,
\newblock Phys. Rev. Lett. 70 (1993) 2491.

\bibitem{JuPo82}
B.~R. Judd and D.~R. Pooler,
\newblock J. Phys. C 15 (1982) 591.

\bibitem{Do89}
M.~C. Downer,
\newblock in {\em Laser Spectroscopy of Solids {II}}, edited by W.~M. Yen,
  page~29, Springer-Verlag, Berlin, 1989.

\bibitem{BuKoRe93}
G.~W. Burdick, H.~J. Kooy, and M.~F. Reid,
\newblock J. Phys.: Condens. Matter 5 (1993) L323.

\bibitem{LiMo82}
I.~Lindgren and J.~Morrison,
\newblock {\em Atomic Many-Body Theory},
\newblock Springer-Verlag, Berlin, 1982.

\bibitem{PiKe75}
M.~S. Pindzola and H.~P. Kelly,
\newblock Phys. Rev. A 11 (1975) 1543.

\bibitem{NgNe85}
B.~Ng and D.~J. Newman,
\newblock J. Chem. Phys. 83 (1985) 1758.

\bibitem{Sm93}
L.~Smentek-Mielczarek,
\newblock Phys. Rev. Lett.  (1993),
\newblock submitted.

\end{thebibliography}

    \newpage

\addtolength{\oddsidemargin} {-1.4cm}

\clearpage
Figure 1: Continued.
\begin{figure}[h]

\begin{picture}( 350.0, 275.0)( -25.0, -50.0)
\thicklines
\put(  50.0, -25.0){ {\Huge (e)}}
\put(  50.0,   0.0){\line(0,1){  50.0}}
\put(  50.0,  30.0){\vector(0,1){0}}
\put(  50.0,  26.0){\vector(0,1){0}}
\put(  55.0,  25.0){ {\Large $i$}}
\put(  50.0,  50.0){\line(1,0){   3.1}}
\multiput(  59.4,  50.0)(  12.5,   0.0){3}{\line(1,0){   6.3}}
\put(  96.9,  50.0){\line(1,0){   3.1}}
\put( 100.0,  50.0){\line(1,1){  10.0}}
\put( 110.0,  60.0){\line(0,-1){  20.0}}
\put( 110.0,  40.0){\line(-1,1){  10.0}}
\put(  50.0,  50.0){\line(0,1){  50.0}}
\put(  50.0,  78.0){\vector(0,1){0}}
\put(  55.0,  75.0){ {\Large $m$}}
\multiput(  54.2, 100.0)(  16.7,   0.0){3}{\oval(   8.3,   8.3)[t]}
\multiput(  62.5, 100.0)(  16.7,   0.0){3}{\oval(   8.3,   8.3)[b]}
\put(  50.0, 100.0){\line(0,1){  50.0}}
\put(  50.0, 128.0){\vector(0,1){0}}
\put(  55.0, 125.0){ {\Large $s$}}
\multiput(  54.2, 150.0)(  16.7,   0.0){3}{\oval(   8.3,   8.3)[t]}
\multiput(  62.5, 150.0)(  16.7,   0.0){3}{\oval(   8.3,   8.3)[b]}
\put(  50.0, 150.0){\line(0,1){  50.0}}
\put(  50.0, 180.0){\vector(0,1){0}}
\put(  50.0, 176.0){\vector(0,1){0}}
\put(  55.0, 175.0){ {\Large $k$}}
\put( 250.0, -25.0){ {\Huge (f)}}
\put( 250.0,   0.0){\line(0,1){  50.0}}
\put( 250.0,  30.0){\vector(0,1){0}}
\put( 250.0,  26.0){\vector(0,1){0}}
\put( 255.0,  25.0){ {\Large $i$}}
\multiput( 254.2,  50.0)(  16.7,   0.0){3}{\oval(   8.3,   8.3)[t]}
\multiput( 262.5,  50.0)(  16.7,   0.0){3}{\oval(   8.3,   8.3)[b]}
\put( 250.0,  50.0){\line(0,1){  50.0}}
\put( 250.0,  78.0){\vector(0,1){0}}
\put( 255.0,  75.0){ {\Large $s$}}
\multiput( 254.2, 100.0)(  16.7,   0.0){3}{\oval(   8.3,   8.3)[t]}
\multiput( 262.5, 100.0)(  16.7,   0.0){3}{\oval(   8.3,   8.3)[b]}
\put( 250.0, 100.0){\line(0,1){  50.0}}
\put( 250.0, 128.0){\vector(0,1){0}}
\put( 255.0, 125.0){ {\Large $m$}}
\put( 250.0, 150.0){\line(1,0){   3.1}}
\multiput( 259.4, 150.0)(  12.5,   0.0){3}{\line(1,0){   6.3}}
\put( 296.9, 150.0){\line(1,0){   3.1}}
\put( 300.0, 150.0){\line(1,1){  10.0}}
\put( 310.0, 160.0){\line(0,-1){  20.0}}
\put( 310.0, 140.0){\line(-1,1){  10.0}}
\put( 250.0, 150.0){\line(0,1){  50.0}}
\put( 250.0, 180.0){\vector(0,1){0}}
\put( 250.0, 176.0){\vector(0,1){0}}
\put( 255.0, 175.0){ {\Large $k$}}

\end{picture}

\end{figure}

\clearpage


 \begin{figure}[h]
 \caption{Some third-order two-electron two-photon absorption diagrams.
Notation is the same as Fig.\ \protect\ref{fig:tpa1}. The Coulomb
interaction  is represented by a dotted line. Folded diagram (e)
dominates the correlation calculation of Ref.\ \protect\cite{BuRe93}.
   \label{fig:tpa2}}

\begin{picture}( 500.0, 550.0)( -25.0,-325.0)
\thicklines
\put(   7.5, -25.0){ {\Huge (a)}}
\put(  50.0,   0.0){\line(0,1){  50.0}}
\put(  50.0,  30.0){\vector(0,1){0}}
\put(  50.0,  26.0){\vector(0,1){0}}
\multiput(  54.2,  50.0)(  16.7,   0.0){3}{\oval(   8.3,   8.3)[t]}
\multiput(  62.5,  50.0)(  16.7,   0.0){3}{\oval(   8.3,   8.3)[b]}
\put(  50.0,  50.0){\line(0,1){  50.0}}
\put(  50.0,  78.0){\vector(0,1){0}}
\put(  50.0, 100.0){\line(-1,0){   3.1}}
\multiput(  40.6, 100.0)( -12.5,   0.0){3}{\line(-1,0){   6.3}}
\put(   3.1, 100.0){\line(-1,0){   3.1}}
\put(  50.0, 100.0){\line(0,1){  50.0}}
\put(  50.0, 128.0){\vector(0,1){0}}
\multiput(  54.2, 150.0)(  16.7,   0.0){3}{\oval(   8.3,   8.3)[t]}
\multiput(  62.5, 150.0)(  16.7,   0.0){3}{\oval(   8.3,   8.3)[b]}
\put(  50.0, 150.0){\line(0,1){  50.0}}
\put(  50.0, 180.0){\vector(0,1){0}}
\put(  50.0, 176.0){\vector(0,1){0}}
\put(   0.0,   0.0){\line(0,1){  50.0}}
\put(   0.0,  30.0){\vector(0,1){0}}
\put(   0.0,  26.0){\vector(0,1){0}}
\put(   0.0,  50.0){\line(0,1){  50.0}}
\put(   0.0, 100.0){\line(0,1){  50.0}}
\put(   0.0, 150.0){\line(0,1){  50.0}}
\put(   0.0, 180.0){\vector(0,1){0}}
\put(   0.0, 176.0){\vector(0,1){0}}
\put( 157.5, -25.0){ {\Huge (b)}}
\put( 200.0,   0.0){\line(0,1){  50.0}}
\put( 200.0,  30.0){\vector(0,1){0}}
\put( 200.0,  26.0){\vector(0,1){0}}
\put( 200.0,  50.0){\line(-1,0){   3.1}}
\multiput( 190.6,  50.0)( -12.5,   0.0){3}{\line(-1,0){   6.3}}
\put( 153.1,  50.0){\line(-1,0){   3.1}}
\put( 200.0,  50.0){\line(0,1){  50.0}}
\put( 200.0,  78.0){\vector(0,1){0}}
\multiput( 204.2, 100.0)(  16.7,   0.0){3}{\oval(   8.3,   8.3)[t]}
\multiput( 212.5, 100.0)(  16.7,   0.0){3}{\oval(   8.3,   8.3)[b]}
\put( 200.0, 100.0){\line(0,1){  50.0}}
\put( 200.0, 128.0){\vector(0,1){0}}
\multiput( 204.2, 150.0)(  16.7,   0.0){3}{\oval(   8.3,   8.3)[t]}
\multiput( 212.5, 150.0)(  16.7,   0.0){3}{\oval(   8.3,   8.3)[b]}
\put( 200.0, 150.0){\line(0,1){  50.0}}
\put( 200.0, 180.0){\vector(0,1){0}}
\put( 200.0, 176.0){\vector(0,1){0}}
\put( 150.0,   0.0){\line(0,1){  50.0}}
\put( 150.0,  30.0){\vector(0,1){0}}
\put( 150.0,  26.0){\vector(0,1){0}}
\put( 150.0,  50.0){\line(0,1){  50.0}}
\put( 150.0, 100.0){\line(0,1){  50.0}}
\put( 150.0, 130.0){\vector(0,1){0}}
\put( 150.0, 126.0){\vector(0,1){0}}
\put( 150.0, 150.0){\line(0,1){  50.0}}

\put( 357.5, -25.0){ {\Huge (c)}}
\put( 400.0,   0.0){\line(0,1){  50.0}}
\put( 400.0,  30.0){\vector(0,1){0}}
\put( 400.0,  26.0){\vector(0,1){0}}
\multiput( 404.2,  50.0)(  16.7,   0.0){3}{\oval(   8.3,   8.3)[t]}
\multiput( 412.5,  50.0)(  16.7,   0.0){3}{\oval(   8.3,   8.3)[b]}
\put( 400.0,  50.0){\line(0,1){  50.0}}
\put( 400.0,  78.0){\vector(0,1){0}}
\put( 400.0, 100.0){\line(-1,0){   3.1}}
\multiput( 390.6, 100.0)( -12.5,   0.0){3}{\line(-1,0){   6.3}}
\put( 353.1, 100.0){\line(-1,0){   3.1}}
\put( 400.0, 100.0){\line(0,1){  50.0}}
\put( 400.0, 150.0){\line(0,1){  50.0}}
\put( 400.0, 180.0){\vector(0,1){0}}
\put( 400.0, 176.0){\vector(0,1){0}}
\put( 350.0,   0.0){\line(0,1){  50.0}}
\put( 350.0,  30.0){\vector(0,1){0}}
\put( 350.0,  26.0){\vector(0,1){0}}
\put( 350.0,  50.0){\line(0,1){  50.0}}
\put( 350.0, 100.0){\line(0,1){  50.0}}
\put( 350.0, 128.0){\vector(0,1){0}}
\multiput( 345.8, 150.0)( -16.7,   0.0){3}{\oval(   8.3,   8.3)[t]}
\multiput( 337.5, 150.0)( -16.7,   0.0){3}{\oval(   8.3,   8.3)[b]}
\put( 350.0, 150.0){\line(0,1){  50.0}}
\put( 350.0, 180.0){\vector(0,1){0}}
\put( 350.0, 176.0){\vector(0,1){0}}
\put(  57.5,-300.0){ {\Huge (d)}}
\put( 100.0,-275.0){\line(0,1){  50.0}}
\put( 100.0,-245.0){\vector(0,1){0}}
\put( 100.0,-249.0){\vector(0,1){0}}
\put( 100.0,-225.0){\line(-1,0){   3.1}}
\multiput(  90.6,-225.0)( -12.5,   0.0){3}{\line(-1,0){   6.3}}
\put(  53.1,-225.0){\line(-1,0){   3.1}}
\put( 100.0,-225.0){\line(0,1){  50.0}}
\put( 100.0,-197.0){\vector(0,1){0}}
\multiput( 104.2,-175.0)(  16.7,   0.0){3}{\oval(   8.3,   8.3)[t]}
\multiput( 112.5,-175.0)(  16.7,   0.0){3}{\oval(   8.3,   8.3)[b]}
\put( 100.0,-175.0){\line(0,1){  50.0}}
\put( 100.0,-125.0){\line(0,1){  50.0}}
\put( 100.0, -95.0){\vector(0,1){0}}
\put( 100.0, -99.0){\vector(0,1){0}}
\put(  50.0,-275.0){\line(0,1){  50.0}}
\put(  50.0,-245.0){\vector(0,1){0}}
\put(  50.0,-249.0){\vector(0,1){0}}
\put(  50.0,-225.0){\line(0,1){  50.0}}
\put(  50.0,-175.0){\line(0,1){  50.0}}
\put(  50.0,-147.0){\vector(0,1){0}}
\multiput(  45.8,-125.0)( -16.7,   0.0){3}{\oval(   8.3,   8.3)[t]}
\multiput(  37.5,-125.0)( -16.7,   0.0){3}{\oval(   8.3,   8.3)[b]}
\put(  50.0,-125.0){\line(0,1){  50.0}}
\put(  50.0, -95.0){\vector(0,1){0}}
\put(  50.0, -99.0){\vector(0,1){0}}

\put( 257.5,-300.0){ {\Huge (e)}}
\put( 250.0,-275.0){\line(0,1){  25.0}}
\put( 250.0,-250.0){\line(0,1){  50.0}}
\put( 250.0,-220.0){\vector(0,1){0}}
\put( 250.0,-224.0){\vector(0,1){0}}
\put( 250.0,-200.0){\line(0,1){  25.0}}
\put( 250.0,-175.0){\line(1,0){   3.1}}
\multiput( 259.4,-175.0)(  12.5,   0.0){3}{\line(1,0){   6.3}}
\put( 296.9,-175.0){\line(1,0){   3.1}}
\put( 250.0,-175.0){\line(0,1){  25.0}}
\put( 250.0,-150.0){\line(0,1){  50.0}}
\put( 250.0,-120.0){\vector(0,1){0}}
\put( 250.0,-124.0){\vector(0,1){0}}
\put( 250.0,-100.0){\line(0,1){  25.0}}
\put( 300.0,-275.0){\line(0,1){  25.0}}
\put( 300.0,-250.0){\line(0,1){  50.0}}
\put( 300.0,-220.0){\vector(0,1){0}}
\put( 300.0,-224.0){\vector(0,1){0}}
\put( 300.0,-200.0){\line(0,1){  25.0}}
\put( 300.0,-175.0){\line(1,-1){  50.0}}
\put( 330.0,-205.0){\vector(1,-1){0}}
\put( 326.0,-201.0){\vector(1,-1){0}}
\put( 325.0,-200.0){\circle{  15.0}}
\multiput( 354.2,-225.0)(  16.7,   0.0){3}{\oval(   8.3,   8.3)[t]}
\multiput( 362.5,-225.0)(  16.7,   0.0){3}{\oval(   8.3,   8.3)[b]}
\put( 350.0,-225.0){\line(0,1){  12.5}}
\put( 350.0,-212.5){\line(0,1){  50.0}}
\put( 350.0,-184.5){\vector(0,1){0}}
\put( 350.0,-162.5){\line(0,1){  12.5}}
\multiput( 354.2,-150.0)(  16.7,   0.0){3}{\oval(   8.3,   8.3)[t]}
\multiput( 362.5,-150.0)(  16.7,   0.0){3}{\oval(   8.3,   8.3)[b]}
\put( 350.0,-150.0){\line(0,1){  12.5}}
\put( 350.0,-137.5){\line(0,1){  50.0}}
\put( 350.0,-107.5){\vector(0,1){0}}
\put( 350.0,-111.5){\vector(0,1){0}}
\put( 350.0, -87.5){\line(0,1){  12.5}}

\end{picture}

 \end{figure}
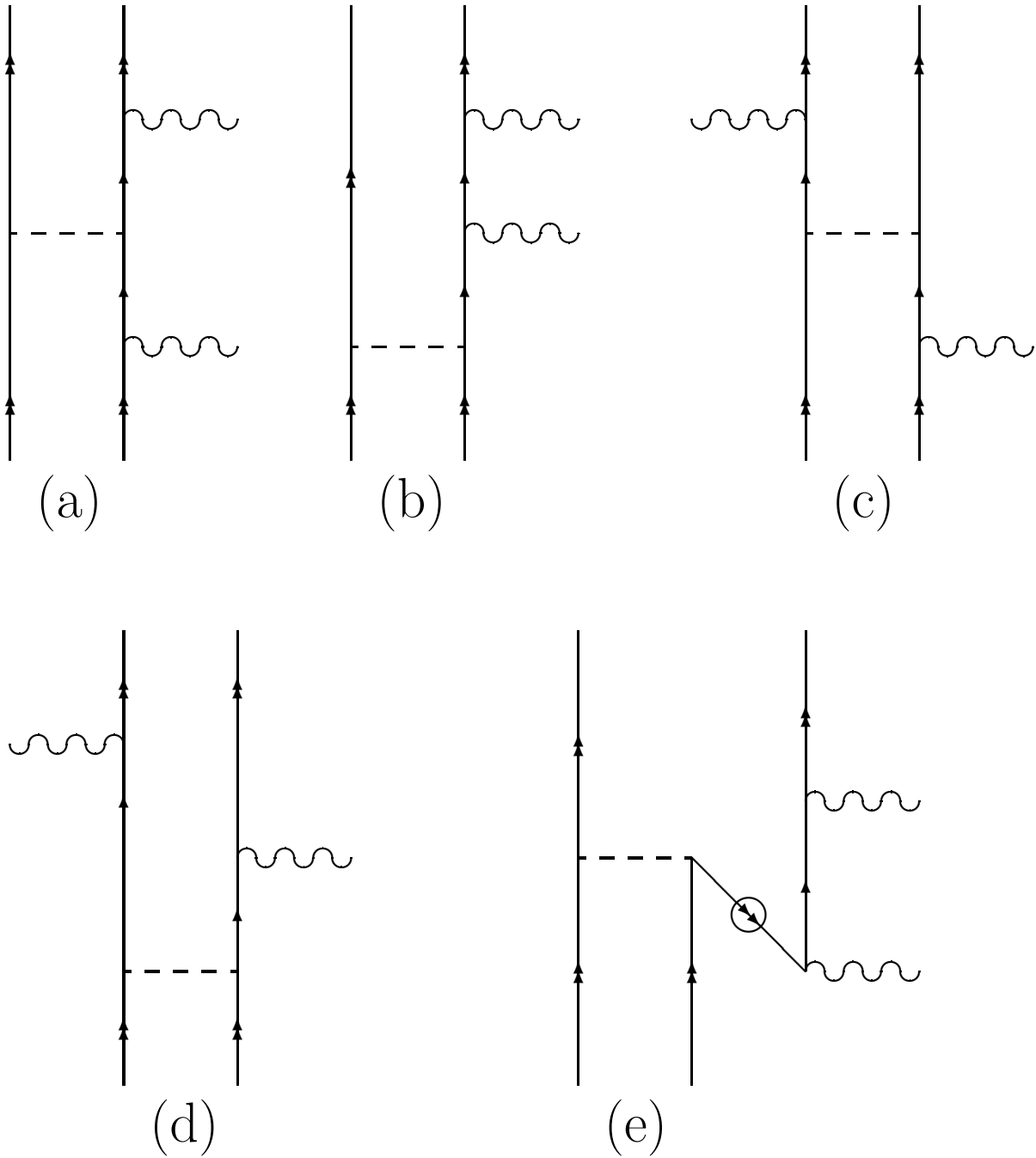

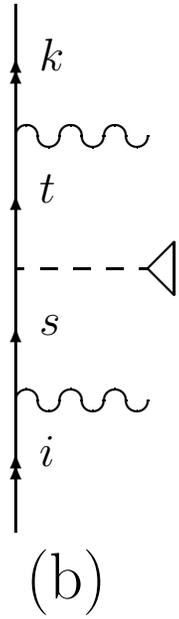
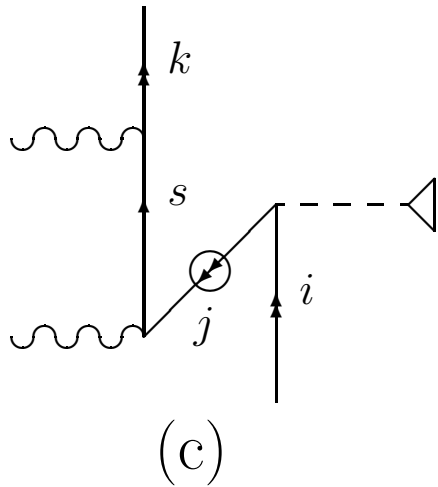
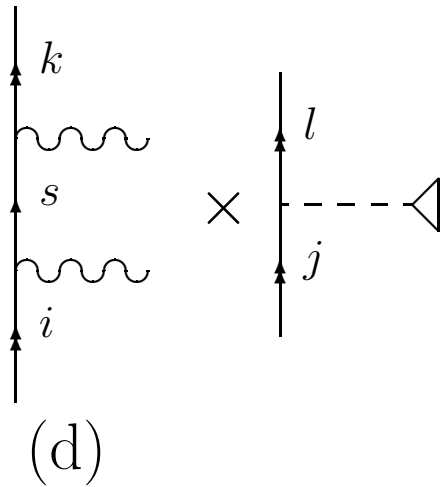
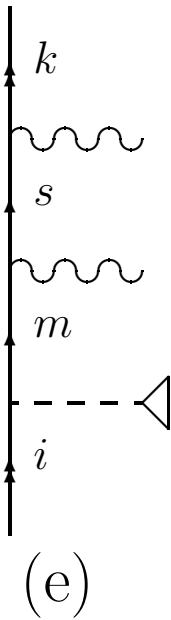
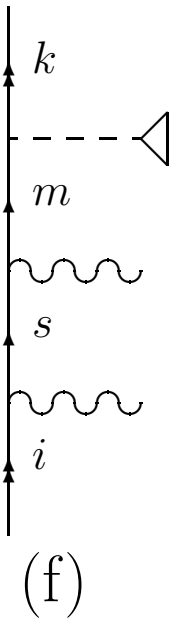
\end{document}